\pdfoutput=1
\documentclass[11pt]{article}
\usepackage{multirow}
\usepackage[table]{xcolor}
\usepackage{amsmath}
\usepackage{acl}

\usepackage{times}
\usepackage{latexsym}
\usepackage{graphicx}
\usepackage{booktabs}

\usepackage[T1]{fontenc}
\usepackage[utf8]{inputenc}

\usepackage{microtype}

\usepackage{inconsolata}

\title{Synchronized Video Storytelling: \\Generating Video Narrations with Structured Storyline}

\makeatletter
\def\thanks#1{\protected@xdef\@thanks{\@thanks
        \protect\footnotetext{#1}}}
\makeatother
\author{Dingyi Yang$^1$, Chunru Zhan$^2$, Ziheng Wang$^1$, Biao Wang$^2$, \bf Tiezheng Ge$^2$, \bf Bo Zheng$^2$, \bf Qin Jin$^1$*\thanks{ *Corresponding Author.}\\
         $^1$Renmin University of China \\ 
         $^2$Alibaba Group \\ 
         \texttt{\{yangdingyi,zihengwang,qjin\}@ruc.edu.cn} \\
         \texttt{\{zhanchunru.zcr,eric.wb,tiezheng.gtz,bozheng\}@alibaba-inc.com} \\
         }

\begin{document}
\maketitle
\begin{abstract}
Video storytelling is engaging multimedia content that utilizes video and its accompanying narration to attract the audience, where a key challenge is creating narrations for recorded visual scenes. 
%This involves understanding sequential visual scenes and crafting coherent narrations to describe the video. 
Previous studies on dense video captioning and video story generation have made some progress. However, in practical applications, we typically require synchronized narrations for ongoing visual scenes. In this work, we introduce a new task of Synchronized Video Storytelling, which aims to generate synchronous and informative narrations for videos. These narrations, associated with each video clip, should relate to the visual content, integrate relevant knowledge, and have an appropriate word count corresponding to the clip's duration. Specifically, a structured storyline is beneficial to guide the generation process, ensuring coherence and integrity.
%In this study, we address this challenge named synchronized video storytelling. It aims to generate coherent stories for videos composed of sequential video clips. Each clip is accompanied by a narrative that describes the visual scene, incorporates prior knowledge, and has an appropriate word length aligned with the clip's duration. Existing works on dense video captioning or video storytelling only generate summaries for video clips, unable to incorporate external knowledge to generate a coherent synchronized narrative for ongoing video. Current Multimodal Large Language Models are not capable of generating narratives for sequential input videos. Generating descriptions for video shots and then applying powerful LLMs in zero-shot or few-shot generation also face challenges in achieving satisfactory results.
To support the exploration of this task, we introduce a new benchmark dataset E-SyncVidStory with rich annotations. 
Since existing Multimodal LLMs are not effective in addressing this task in one-shot or few-shot settings, we propose a framework named VideoNarrator that can generate a storyline for input videos and simultaneously generate narrations with the guidance of the generated or predefined storyline. We further introduce a set of evaluation metrics to thoroughly assess the generation. Both automatic and human evaluations validate the effectiveness of our approach. Our dataset, codes, and evaluations are released at \url{https://github.com/alibaba/alimama-video-narrator}.

%can generate a structured storyline based on the input video. This framework can either use the generated storyline or a user-offered one to guide the story-generation process.  We introduce a set of evaluation metrics to thoroughly assess the generated results. Both automatic and human evaluations validate the effectiveness of our approach.
\end{abstract}

\section{Introduction}\label{sec:intro}

%\textbf{[Background] }

%- Introduce video storytelling ;

%- Summarize the limitations of existing video storytelling/captioning tasks; 

%- Introduce controllable (?) video storytelling.

%Video storytelling is a tactic that utilizes the form of video to tell a story. It is widely used as a powerful marketing tool, for educational purposes, or for entertainment. Recently, there has been a focus on generating videos based on pre-existing written stories \cite{he2023animate, zhou2022magicvideo}. Another important situation to note is that individuals have recorded some video scenes. By integrating synchronized narratives, these initial recordings can be transformed into compelling video stories.
Video storytelling \cite{li2019video,bhattacharya2023video} is a tactic that utilizes the naturally engaging video format to share stories and has been widely used as a powerful tool for marketing, educational, or entertainment purposes. The video and its accompanying narration are both critical to successful video storytelling. 
%Therefore, generating/retrieving videos based on written stories \cite{he2023animate, zhou2022magicvideo} or generating narrations based on videos \cite{li2019video} have been two sides explored for the automatic generation of video storytelling. 
Therefore, two aspects of automatic video storytelling have been explored: generating or retrieving videos to illustrate written stories \cite{he2023animate, lu2023show}; and generating narrations based on visual scenes \cite{li2019video}.
%As people often make visual recordings very conveniently nowadays, these initial recordings can be transformed into compelling video stories by integrating synchronized narrations. 
Nowadays, people frequently use convenient devices to make visual recordings. These initial recordings can be transformed into compelling video stories by integrating synchronized narrations. 
There is a strong need for automatically generating narrations for a given video.

\begin{figure}[t]
    \centering
    \includegraphics[width=0.48\textwidth]{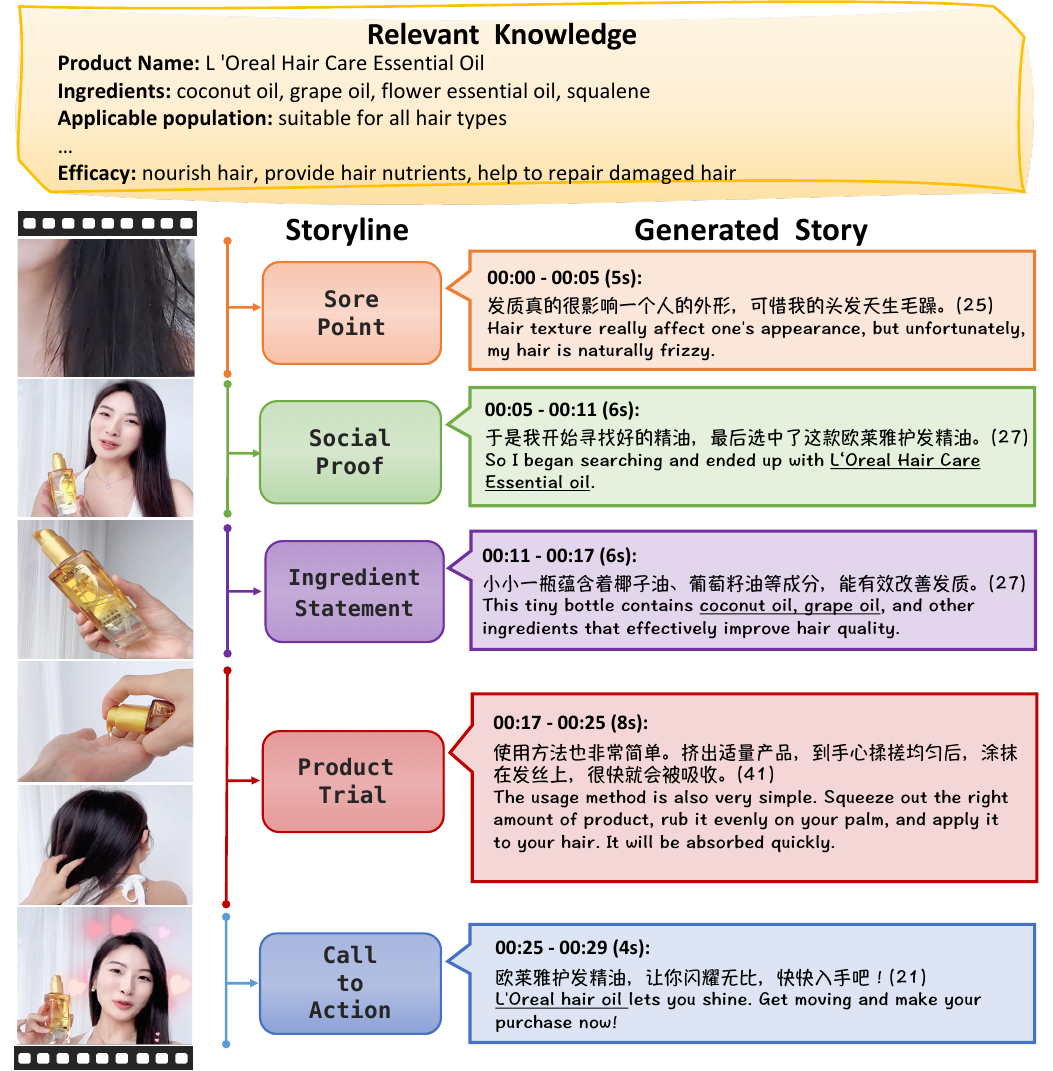}
    \vspace{-8pt}
    \caption{An example of our annotated synchronized video storytelling, with English translations provided for easy reading. Sequential visual scenes are narrated, following a storyline expressed by categorized script labels. Each narration should incorporate appropriate relevant knowledge (as underlined in the figure) and have a word count that fits the duration of each clip. }
    \label{fig:task}
    \vspace{-4pt}
\end{figure}

\begin{figure*}[t]
    \centering
    \includegraphics[width=1\textwidth]{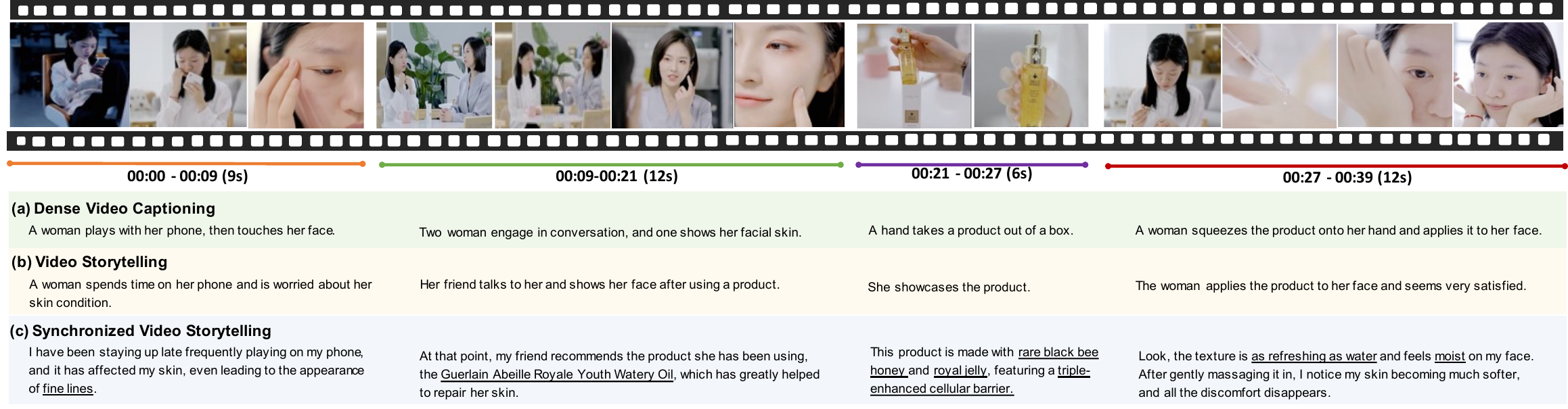}
    \vspace{-8pt}
    \caption{Comparison of the proposed Synchronized Video Storytelling and existing video-to-text generation tasks. (a) Dense Video Captioning aims to generate corresponding narrations for multiple events within a video. (b) Video Storytelling aims to form a coherent story with these narratives. (c) The proposed Synchronized Video Storytelling aims to generate synchronized, informative, and coherent narrations to support voiceovers for videos.} %Compared to Video Storytelling, each narration should fit the event duration, and incorporate relevant knowledge. }
    \label{fig:task_compare}
    \vspace{-2pt}
\end{figure*}
%\textbf{[Existing methods]}

%- The challenges of our task. Traditional benchmarks and methods are unable to solve.

%- The efficiency of LLMs and LMMs. However:

%- - Problem of LMMs: (1) hard to achieve time alignment; (2) Poor controllable ability; (3)Long visual tokens; (4) Severe hallucinations

%- - Problem of pipelines (Multi-modal+LLM): (1) Rely on the efficiency of captioning model \& Error accumulation; (2)Unable to solve controllable problem

Existing research on video-to-text generation mostly focuses on creating a single sentence to summarize the video \cite{wang2019vatex} or generating fine-grained captions for video clips \cite{krishna2017dense}. \citet{gella2018dataset} and \citet{li2019video} take a step further by generating sequential visual summaries for video scenes, creating a coherent video story. However, in real-life applications, when narrating an ongoing video, each narration should not only reflect the current visual scene, but also have an appropriate word sequence length to fit the duration of the video clip. Including relevant external knowledge is also important, as it can attract and convince the audience, particularly for marketing purposes.
Additionally, we believe that a logical storyline helps maintain the coherence and integrity of the narratives.

In this paper, we introduce the task of \textbf{Synchronized Video Storytelling}, which aims to generate synchronized narrations for videos. As illustrated in Figure \ref{fig:task}, the sequential visual scenes in the video are accompanied by corresponding narrations. Each narration has a suitable word sequence length and incorporates relevant knowledge to effectively engage the audience. They can be directly utilized as voiceovers for video recordings. This distinguishes our task from existing ones, as illustrated in Figure \ref{fig:task_compare}. Furthermore, our generation follows a structured storyline, similar to the human writing process, beginning with a brief outline and then drafting the complete story \cite{yang2022doc}. %As displayed in Figure , the main difference is that the coherent narrations contain more compelling information and are of an appropriate length, which can directly be used as voiceovers for the video recordings.% We compare the task with existing video-to-text generation tasks in Figure \ref{fig:task_compare}. 
 This proposed task is complex and cannot be effectively solved by traditional frameworks or evaluated by existing benchmarks.  A possible approach is to utilize powerful Multi-modal Large Language Models (MLLMs) \cite{han2023imagebind,sun2023generative} for zero-shot or few-shot generation. However, most MLLMs face challenges with time alignment and sequential video comprehension \cite{huang2023vtimellm}. Moreover, generating narrations for longer videos becomes difficult for MLLMs due to the requirement of excessively long visual tokens.
Another approach is to generate captions for video shots and use them as input for the LLMs \cite{bhattacharya2023video, lin2023mm}. However, this relies heavily on the performance of the captioning model. Furthermore, current models are ineffective in generating coherent, appealing, and length-appropriate narrations. Additional training on specific instruction datasets is necessary.

To support the exploration of the proposed task, we create a benchmark dataset named E-SyncVidStory (\textbf{E-}commerce \textbf{Sync}hronized \textbf{Vid}eo \textbf{Story}telling). Given the rapid growth of e-commerce platforms, many sellers have product information and visual materials, but may lack the time and resources to convert these into engaging advertisement videos. Therefore, they seek an automatic and efficient solution. Our dataset can provide value for both research and practical applications in this area. E-SyncVidStory contains 6.0K videos with 41.3K video clips in total. Each video is segmented as sequential visual scenes with synchronized Chinese narrations. As the primary objective of Ad stories is to engage with the audience and showcase products \cite{bhattacharya2023video}, they might not always present a sequence of events as a classic story. Still, like classic ones, advertising stories should be coherent and follow a logical storyline, which is annotated in our dataset\footnote{The storyline is expressed by categorized script
labels. These labels are divided into 12 types, with their definitions provided in Appendix \ref{sec:scripts}.}. All annotations are obtained through automatic preprocessing, GPT-based refinement, and manual checks.
%We collect high-quality advertisement videos with informative content and a high click-through rate. These videos are segmented as sequential visual scenes with corresponding narrations. Each narration in a story follows a storyline that is represented by a combination of some specific script labels\footnote{The script labels are categorized into 12 types, and their definitions are displayed in Appendix \ref{sec:appendix}.}.%, which together construct the entire storyline. 

We further propose an effective framework integrating vision models and LLMs to address the task of synchronized video storytelling. % All of the information is constructed with a specifically designed instruction structure, which simultaneously conducts the problem of storyline generation and controllable story generation (controlled by the storyline and word length range). 
Our specifically designed instruction structure allows simultaneous generation of the storyline and the synchronized story. To enhance the effectiveness of visual embedding, we compress the long visual tokens into shorter visual information, which includes the long-term memory of previous clips and compressed information of the current clip. To evaluate the generated stories, we propose a set of evaluation metrics to comprehensively measure the generated results. % on E-SyncVidStory. We also adapt the Video Storytelling dataset \cite{li2019video} to fit our task. Following this, we carry out experiments on this general domain dataset to confirm the efficiency of our proposed framework.
Extensive experiments on both specific domain (E-SyncVidStory) and general domain \cite{li2019video} Video Storytelling verify the effectiveness of our proposed framework.

%We collect high-quality advertisement videos with informative content and a high clicking rate, these videos are automatically segmented into video clips that have coherent and attractive narratives. All the results are then manually refined. Specifically, we summarize a structured storyline for each story. With the guidance of a logical storyline, the generated stories can become more coherent and complete. For example, as the case displayed in Figure \ref{fig:task}, although the second shot and the last shot have visually similar scenes, the last one is more likely to evoke a response from the audience rather than introduce the product, in order to conclude the whole story. 

%\textbf{[Introduce our Contribution]}
%- Introduce a benchmark

%- effective framework

%- effective prompting design

%- Visual shot encoding

The main contributions of our work include:
%\begin{itemize}
1) We introduce the new task of synchronized video storytelling, which is more challenging as it requires generating a synchronous, informative, and coherent story for a given video. %guided by external knowledge and a structured storyline. To the best of our knowledge, it is the first work to address this practical and challenging problem; 
2) To support the exploration of this task, we collect a benchmark dataset in the advertising domain, namely E-SyncVidStory, which contains rich annotations, and also can support more research beyond video storytelling; 
3) We propose an effective framework called VideoNarrator, which takes relevant knowledge and sequential video scenes as inputs. It simultaneously supports storyline generation and controllable video story generation; 
4) We introduce a set of systematic evaluation metrics to comprehensively measure the story generation results. Automatic and human evaluation results verify the effectiveness of our method.
%\end{itemize}      
\vspace{-2px}
\section{Related Works}
\vspace{-2px}
\iffalse
\begin{table*}[t]
\centering
\fontsize{7.3}{17}\selectfont
\caption{\label{table:dataset}Comparison of our proposed dataset and existing dense video captioning and video storytelling dataset.}
\begin{tabular}{l|l|cccccc}
\toprule
 & \textbf{Domain} & \textbf{Num. videos} & \textbf{\begin{tabular}[c]{@{}c@{}}Avg. clips\\ (per video)\end{tabular}} & \textbf{Avg. video len.} &  \textbf{\begin{tabular}[c]{@{}c@{}}Avg. text len.\\ (per video)\end{tabular}} & \multicolumn{1}{l}{\begin{tabular}[c]{@{}l@{}}\textbf{Avg. text len.}\\ \textbf{(per second)}\end{tabular}} & \textbf{Storyline} \\
 \midrule
YouCook2 & cooking & 2K & 7.7 & 320s & 177.4 (en) & 0.55 & - \\
ViTT & cooking / instructional & 8K & 5  & 250s & 110.5 (en) & 0.44 & - \\
ActivityNet Captions & human activities & 20K & 3.7 & 120s & 52.5 (en) & 0.44 & - \\
Charades-STA & indoors activities & 10K & - & 30s  & 6.07 (en) & 0.20 & - \\
\midrule
Video Storytelling & open & 105 & 13.5 & 755s& 162.6 (en) & 0.22 & - \\
Ours & e-commerce & 6K & 7.3& 38s & 203.85 (zh) & 5.17 & yes\\
\bottomrule
\end{tabular}
\end{table*}
\fi

\paragraph{Video Narration Generation.}
Bridging vision and natural language is a longstanding goal in computer vision and multimedia research \cite{li2019video}. Previous research on video-to-text generation primarily focuses on video captioning tasks, generating single-sentence factual descriptions for the whole video \cite{wang2019vatex, xu2016msr}. Some works aim to provide more detailed and comprehensive descriptions by generating multi-sentence paragraphs \cite{krishna2017dense, zhou2018towards}. However, most of these works ignore the overall coherence of the video descriptions. \citet{gella2018dataset} and \citet{li2019video} step further to generate a coherent and succinct story from abundant and complex video data. However, these datasets only generate short summaries for long videos and do not handle synchronized video storytelling. Additionally, all these works only focus on the visual content and do not take into account the need to incorporate external knowledge.

\paragraph{Multi-modal Large Language Models.}
With the success of Large Language Models (LLMs), many works have tried to build Multi-modal LLMs by combining models from other modalities. In the video field, VideoChat \cite{li2023videochat} integrates video foundation models and LLMs via a learnable neural interface, excelling in spatiotemporal reasoning, event localization, and causal relationship inference. VideoChat Longvideo \cite{opengv23longvideo} incorporates LangChain into VideoChat to support videos that are longer than one minute. Video-ChatGPT \cite{maaz2023video_chatgpt} design a 100k video instruction dataset, successfully empowering LLMs to comprehend videos. Video-LLaVA \cite{lin2023videollava} learns from a mixed dataset of images and videos, and performs joint visual reasoning on images and videos. However, all of these methods are not effective in comprehending sequential video clips. VtimeLLM \cite{huang2023vtimellm} design a boundary-aware three-stage training strategy, enhancing results for dense video comprehension. However, it struggles with longer video inputs since the maximum training length is 2048. Like other MLLMs, it also lacks efficiency in generating video descriptions that are constrained by appropriate text length and specific storylines.

\iffalse
\begin{figure}[t]
    \centering
    \includegraphics[width=0.8\linewidth]{latex/images/frame_num.pdf}
    \caption{Distribution of clip numbers per video.}
    \label{fig:frame_num}
\end{figure}

\begin{figure}[t]
    \centering
    \includegraphics[width=0.8\linewidth]{latex/images/framelength.pdf}
    \caption{Distribution of the duration of the video clips.}
    \label{fig:clip_length}
\end{figure}
\fi

\section{Synchronized Video Storytelling}
\vspace{-3px}
\subsection{Task Definition}\label{sec:task_define}
\vspace{-2px}
\begin{table*}[t]
\centering
\fontsize{6.7}{7.5}\selectfont
\caption{\label{table:dataset} Comparison between E-SyncVidStory and existing dense video captioning (top half) and video story generation (lower half) datasets.}
\vspace{-6pt}
\begin{tabular}{l|c|c|ccccc}
\toprule
 & \textbf{Input Modality} & \textbf{Storyline}  & \textbf{Num. videos} & \textbf{\begin{tabular}[c]{@{}c@{}}Avg. clips\\ (per video)\end{tabular}} & \textbf{\begin{tabular}[c]{@{}c@{}}Avg. text len.\\ (per video)\end{tabular}} & \multicolumn{1}{l}{\begin{tabular}[c]{@{}l@{}}\textbf{Avg. text len.}\\ \textbf{(per second)}\end{tabular}}\\
 \midrule
YouCook2 \cite{zhou2018youcook2} & Video & - & 2K & 7.7 & 67.7 (en) & 0.21 \\
ActivityNet Captions \cite{krishna2017dense} & Video & - & 20K & 3.7 &  47.6 (en) & 0.40 \\
Charades-STA \cite{gao2017tall} & Video & - & 10K & 1.8 & 11.0 (en) & 0.36 \\
ViTT \cite{huang2020multimodal} & Video & - & 8K & 5  & 110.5 (en) & 0.44 \\
\midrule
VideoStory (unreleased) \cite{gella2018dataset} & Video & - & 20K & 6.1 &  - & - \\
Video Storytelling \cite{li2019video} & Video & - & 105 & 13.5 & 162.6 (en) & 0.22 \\
%\cite{bain2020condensed} & Video & - &  &  &  (en) &  \\
%\cite{lu2023show} & Video & - &  &  &  (en) &  \\
\textbf{E-SyncVidStory} & Video \& Knowledge & yes & 6K & 6.9 & 194.1 (zh) & 5.21 \\
\bottomrule
\end{tabular}
\end{table*}

%\paragraph{Generating Synchronized Video Narration} 
Given a video $V$ composed of sequential video clips $\{v_1, v_2, ..., v_n\}$ and the related knowledge composed of several knowledge items $\mathcal{K}=\{k_1,k_2,...,k_m\}$, synchronized video storytelling aims to incorporate the knowledge to generate a narration $s_i$ for each video clip $v_i$, and all narrations form a coherent story $\mathbf{S}=\{s_1, s_2, ..., s_n\}$ for the video. The word count \footnote{If the time length of video clip $v_i$ is $t$ seconds, considering the normal speaking speed of 5 words per second, the appropriate word length range can be $[5t-3, 5t+3]$ with an error margin of 3 words in either direction.} of each $s_i$ should be within the appropriate range  $r_i$. To enhance the coherence of the overall story, the generation is guided by a predefined or automatically predicted storyline, which is expressed using a sequence of script labels $\{l_1, l_2, ..., l_n\}$. Each script label $l_i$, belonging to one of the 12 predefined types, summarizes the key point of the corresponding narration.
See Appendix~\ref{sec:scripts} for script label details.

%\paragraph{Script Labels} Script Labels are predefined structure to control the generated story. In this work, we focus on advertising stories in the e-commercial era. We analyze the collected advertising stories and categorize the script labels into 12 types: sore point, social proof/influence, design of appearance, ingredient/material/texture statement, product trial, product effect, product security, specific characteristics, authoritative certificate, production process, call to action, and others. The detailed definitions of these labels are displayed in \ref{sec:scripts}. Each of them corresponds to one of the advertising persuasion strategies \cite{kumar2023persuasion}. %commonly used in the advertising domain. 

\begin{figure}[t]
    \centering
    \includegraphics[width=0.75\linewidth]{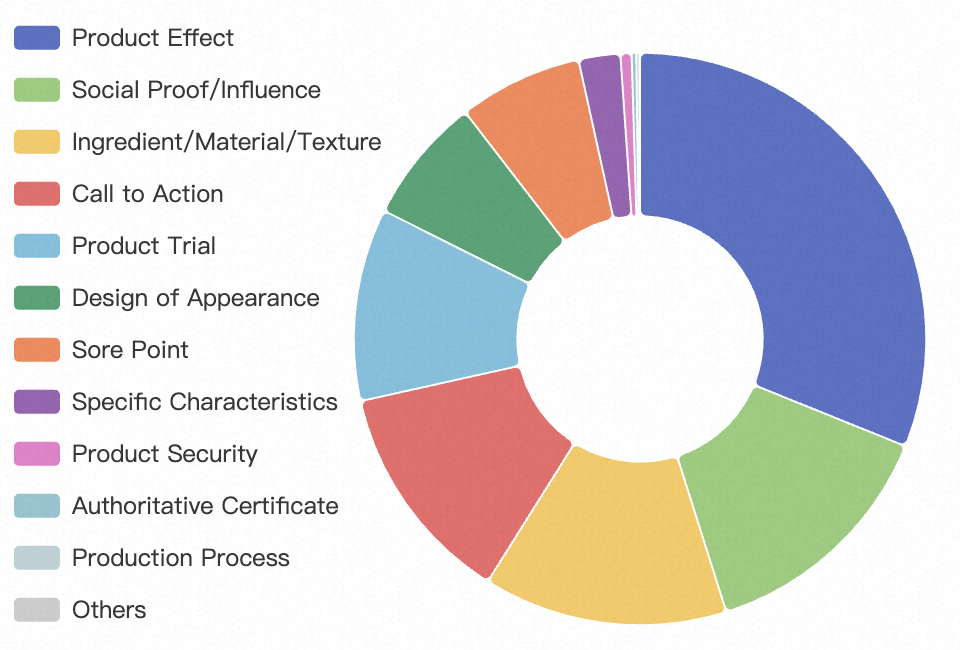}
    \vspace{-8pt}
    \caption{The distribution of the number of 12 types of script labels in the benchmark.}
    \label{fig:script_label}
\end{figure}

\subsection{E-SyncVidStory Dataset}

\paragraph{Data Collection.}
We select high-quality advertisement videos with high click-through rates from the web and collect these videos along with their related information which will be utilized as knowledge points. As shown in Figure \ref{fig:task}, knowledge points are some short expressions about different attributes of the product.
%\qin{elaborate a bit more about knowledge, such as providing some examples? in the footnote? knowledge is used in following evaluation metrics, it will be confusing without proper introduction} 
We automatically divide each video into sequential video clips \cite{mmaction2019} and obtain their synchronized narrations through the automatic speech recognition (ASR) tool \cite{zhang2022paddlespeech}. However, these pre-processed narrations still contain errors. To minimize the labor costs to correct them, we utilize the powerful tool GPT-4 \cite{OpenAI2023GPT4TR} to correct the errors and predict script labels for each narration. The prompts for ASR refinement and script label classification are displayed in our Appendix \ref{sec:prompt}. With the aforementioned automatic pre-processing process, the results are already quite satisfactory. We then recruit crowd workers to review the narrations, check script label classification, and correct any remaining errors. This way, we build our proposed dataset, named \textbf{E-SyncVidStory}.

\begin{figure}[t]
    \centering
    \includegraphics[width=0.73\linewidth]{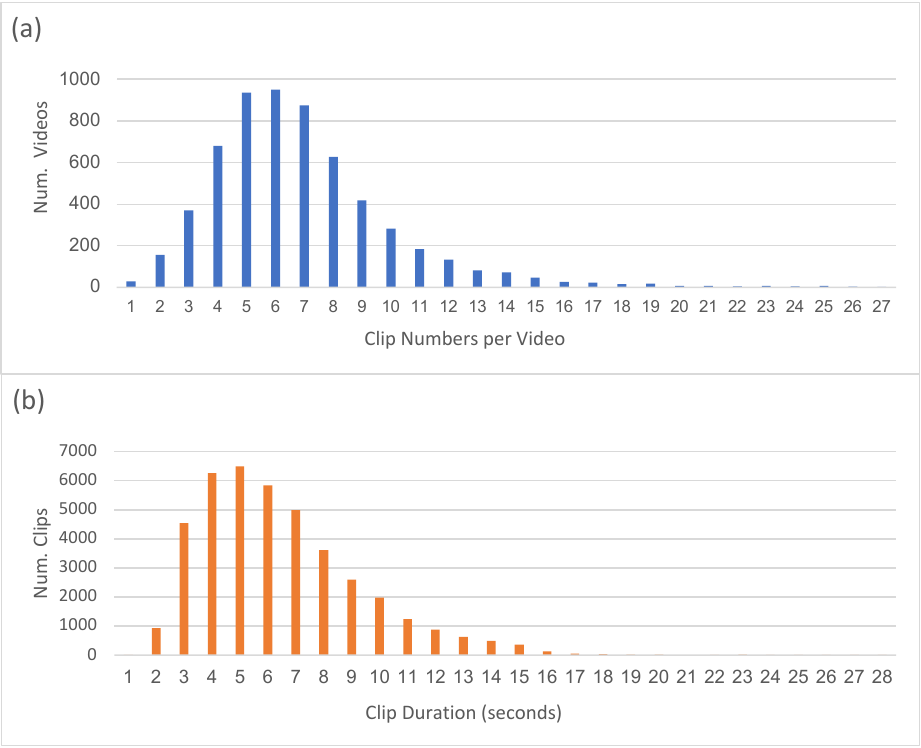}
    \vspace{-8pt}
    \caption{Distribution of clip numbers per video and the clip duration.}
    \label{fig:distribution}
\end{figure}

\paragraph{Statistics and analysis.}
E-SyncVidStory contains 6,032 videos, with a total of 41,292 video clips. The advertising videos cover various industries including personal care, makeup, clothing, household supplies, maternal parenting, and electronics, etc. The video length ranges from 2 to 220 seconds with an average of 39 seconds, and the average story length is 194 words. As shown in Figure \ref{fig:distribution} (a), the clip numbers within each video vary greatly. Thus we apply relative clip position embedding as described in Section \ref{sec:position_embed}. The duration of video clips is displayed in Figure \ref{fig:distribution} (b). Since a single clip within a video can have a maximum duration of 30 seconds, visual compression is needed when comprehending the video. The distributions of script labels are shown in Figure \ref{fig:script_label}. Most narrations focus on the product effect, demonstrating the product's benefits. They also prefer social proof, using positive experiences to persuade the audience. While this work focuses on creating a story from a structured video, this dataset could also aid investigation in generating a well-structured video from multiple unsequential clips, as discussed in Appendix \ref{sec:beyond}.

\paragraph{Dataset Comparison.}
To the best of our knowledge, E-SyncVidStory is the first synchronized video storytelling benchmark. As shown in Table \ref{table:dataset}, the average text length per second is much longer than the previous datasets because it allows for synchronous narration. We also annotate the structured storyline and corresponding knowledge of videos, which can be used to generate more informative and coherent narrations.

\begin{figure*}[t]
    \centering
    \includegraphics[width=0.85\textwidth]{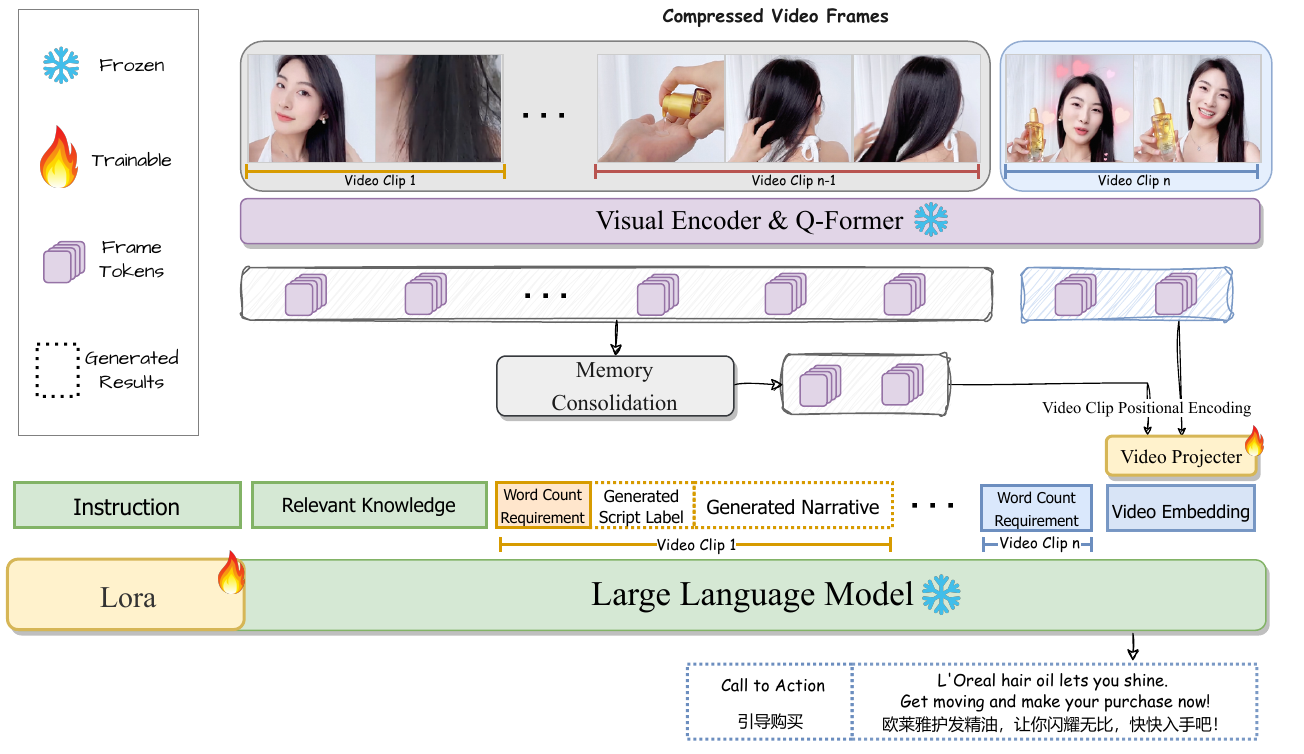}
    \caption{Illustration of our proposed VideoNarrator. The architecture is based on the Visual Feature Extractor, the Video Projector, and the LLM. The original video is compressed, retaining only the important frames. When generating the narration for video clip n, the previous frames are combined into fixed-length video memory tokens and concatenated with the current frame tokens. We pass the concatenated visual features and the relative video clip position embedding through the video projector to achieve the final video embedding. Based on the visual embedding and the previously generated narrations, the LLM will generate an appropriate script label and apply it to guide the narration generation.}
    \label{fig:framework}
\end{figure*}

\subsection{Automatic Evaluation Metrics}\label{sec:metrics}
Following \citet{li2019video}, we evaluate the generated narrations with NLP metrics such as BLEU \cite{papineni2002bleu}, METEOR \cite{banerjee2005meteor}, and CIDEr \cite{vedantam2015cider}. However, it is not reasonable to only consider the scores based on Ground Truth, as the generated story may include different knowledge even from the same aspect. For example, while the visual scene displays the list of ingredients, the narration might select a portion of them to describe. Thus we propose reference-free metrics to evaluate the results, in the aspect of visual relevance, knowledge relevance, controllable accuracy, and fluency.

\paragraph{Visual Relevance.}
Following \citet{DBLP:conf/cvpr/ShiYXYLHZ22}, we use EMScore and $\textrm{EMScore}_{\textrm{ref}}$ to evaluate the visual relevance between the video and generated narrations. The textual and visual embeddings for evaluation are encoded using Chinese-Clip \cite{chinese-clip}. Unlike the traditional visual relevance evaluation, we only consider the similarity between visually relevant words\footnote{We tokenize $s_i$ using jieba \cite{jieba} and retain only the nouns, verbs, and adjectives as visually relevant words.} and the video clips. Because connecting words and knowledge within a sentence which enriches the video narrations should not hurt the visual relevance even if they have a low similarity score.  % have a low similarity score with the visual content, but they do not really affect the visual relevance.

\paragraph{Knowledge Relevance.} \label{sec:knowledge_relevance}
A compelling advertising story should incorporate relevant knowledge to effectively attract customers. Firstly, each narration should incorporate relevant knowledge points while avoiding unrelated messages, denoted as information similarity ($\textrm{Info}_\textrm{Sim}$). Additionally, the knowledge points within a story should be diverse, avoiding the emphasis on only a small amount of information. This is referred to as information diversity ($\textrm{Info}_\textrm{Diverse}$). The detailed evaluation process is displayed in our Appendix \ref{sec:eval_function}.

\paragraph{Controllable Accuracy.}
%We consider the word sequence length accuracy and script label accuracy. An ideal narration $s_i$ for each clip should have an appropriate word sequence length and adhere to the script label $l_i$. The script label accuracy is classified by GPT-4.
We evaluate the accuracy of the word sequence length by checking if the word count of sub-story $s_i$ falls within the appropriate range as described in Section \ref{sec:task_define}. When generating stories controlled by a predefined storyline, we utilize GPT-4 to assess the controllable accuracy of the script labels. The prompt for this evaluation can be found in Appendix \ref{sec:prompt}.

\paragraph{Fluency.} %We apply intra-story repetition \cite{yao2019plan} as a fluency metric, quantifing sentence repetition within a story by measuring word overlaps.
We apply intra-story repetition \cite{yao2019plan} which measures sentence repetition within a story through word overlaps.

\section{Method}

%\subsection{Overview}\label{sec:method_overview}
As shown in Figure \ref{fig:framework}, our proposed VideoNarrator consists of the visual feature extractor, the memory consolidation component, the video projection layer, and the Large Language Model (LLM). Inspired by \citet{maaz2023videohatgpt}, we utilize the powerful CLIP-L/14 visual encoder \cite{radford2021learning} and the Q-Former module from BLIP-2 \cite{li2023blip} to obtain frame-level visual features. These features already achieve a good alignment with the textual modality. To capture temporal information, we further feed them into a linear video projector. However, inputting all video frames would result in significant computational complexity and memory usage \cite{song2023moviechat}. Moreover, it would make it challenging to extract prior knowledge or maintain coherence with the previously generated story. To address these difficulties, we consolidate the visual information from previous clips into fixed-length memory tokens. Additionally, we compress the visual frames of the current clip, retaining only the key information. 

\subsection{Visual Embedding}
\paragraph{Visual Compression.} \label{sec:visual_compress}
Considering the visual information redundancy \cite{tong2022videomae}, we first compress the original frames in each video clip by removing a frame from any adjacent pair frames if their similarity is above a threshold value $\tau$. This removal process will be continued until the similarity between each pair of adjacent frames falls below $\tau$.

%Instead of randomly sampling, we calculate the similarity of adjacent frames and remove one frame in the most similar adjacent pairs. This threshold is repeated until the highest similarity is lower than a threshold value $\tau$.
\paragraph{Memory Consolidation.}
For long videos consisting of multiple video clips, when generating a narration for the current clip $v_i$, we only need a more concise message of previous video clips. Inspired by \citet{song2023moviechat}, we perform memory consolidation by merging the most similar visual tokens. Specifically, we maintain fixed-length tokens as visual memory. We iteratively repeat the merging process \footnote{Assume \{$f_i$\}$_{i=1}^k$ are the visual features of k frames. In each merging process, we
: (1) Evaluate the cosine similarity of each adjacent pair.
(2) Find the pair ($f_m, f_{m+1}$) with the highest similarity. (3)
Merge $f_m$ and $f_{m+1}$ by averaging their features and remove $f_{m+1}$.} until the memory tokens of previous video clips reach the fixed length.

\paragraph{Video Clip Positional Encoding.}\label{sec:position_embed}
The relative position of each video clip is important in video story generation. For example, when approaching the end of a video, it is more likely to evoke a response from the audience rather than introduce the product. However, the model could only acknowledge the absolute position of a video clip. Thus, we propose a relative clip position embedding and add it to the video embeddings. If a clip is located at $p\%$ of the whole video,  then $p$ is its relative position. This value $p$ is passed through a positional embedding layer, whose parameters are updated during training. The encoded position feature is then added to the visual embeddings.

\begin{figure}[t]
    \centering
    \includegraphics[width=0.9\linewidth]{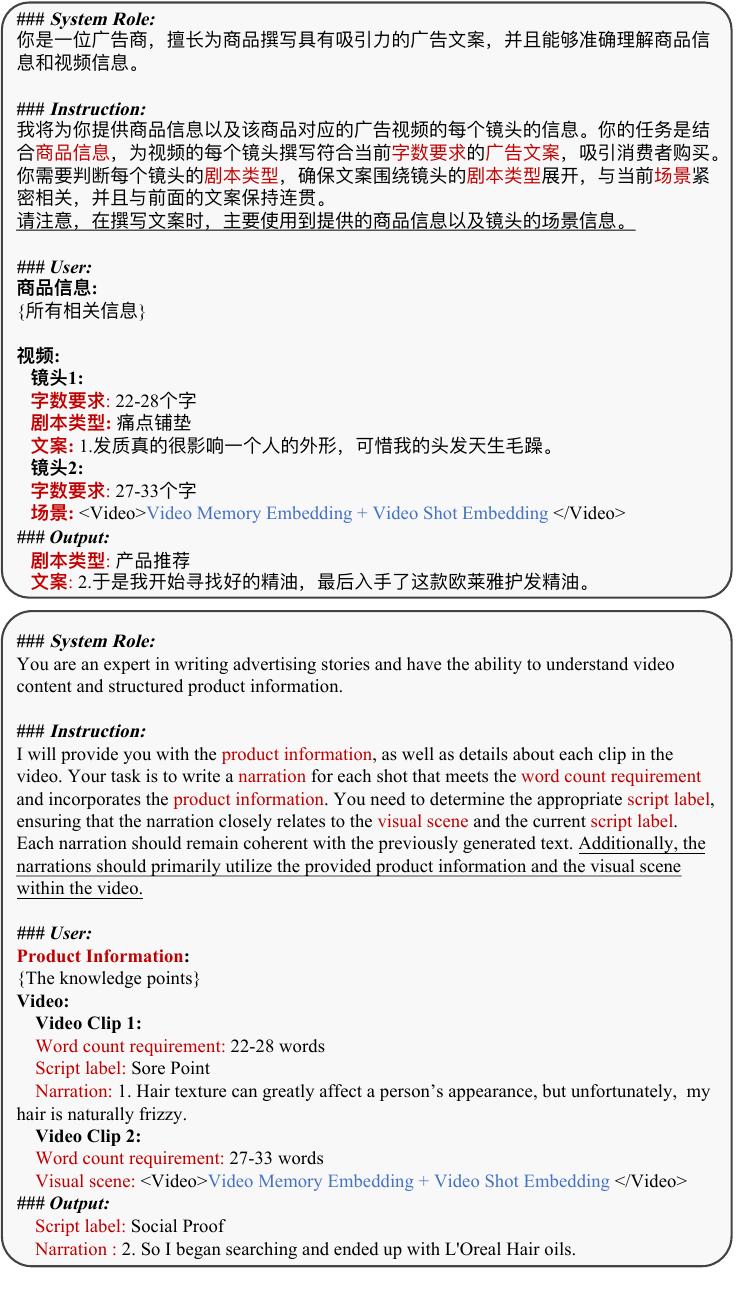}
    \caption{This figure illustrates our system prompt, with English translations provided
for easy reading. The instructions are designed to help the model understand the task of synchronized video storytelling. Here, we present the constructed training sample for the second clip within a video.}
    \label{fig:prompt}
\end{figure}

\begin{table*}[t]
\caption{\label{table:auto_metric} Automatic Evaluation Results on E-SyncVidStory. The best result on each aspect is \textbf{bolded}.}
\vspace{-8pt}
\centering
\fontsize{7.5}{8.2}\selectfont
\begin{tabular}{ll|cccccc}
\toprule
%\textcolor{gray}{
&& \multicolumn{1}{c}{\textbf{CIDEr}} & \textbf{METEOR }  & \multicolumn{1}{c}{\textbf{BLEU-1}} & \multicolumn{1}{c}{\textbf{BLEU-2}} & \multicolumn{1}{c}{\textbf{BLEU-3}} & \multicolumn{1}{c}{\textbf{BLEU-4}} \\ \midrule
 \multirow{2}{*}{Multi-Model Pipelines} & LLaVA-1.5+GPT-3.5 (Zero-Shot) &  15.0& 8.5  & 18.4 & 7.6 & 4.3 & 2.7\\ 
%\multicolumn{1}{l|}{LLaVA1.5+GPT3.5 (One-Shot)} &  21.1& 8.6 & 4.9 & 3.1 & 17.1 & 8.9 \\ 
& LLaVA-1.5+GPT-3.5 (Few-Shot) & 18.6 & 8.3 &  20.0 & 8.5 & 4.9 & 3.2 \\ 
\midrule
\multirow{3}{*}{End2end MLLMs} & Video-ChatGPT \cite{maaz2023video_chatgpt} & 4.1 & 8.6 & 13.8 & 5.5 & 2.9 & 1.7 \\
&Video-LLaVA \cite{lin2023videollava} & 6.1 & 8.3 & 15.1 &6.0 & 3.3  & 1.8\\
&VTimeLLM \cite{huang2023vtimellm}& 8.4 & 8.6 & 14.9  & 6.5 &3.9 &2.6 \\
\midrule
\multirow{7}{*}{\begin{tabular}[c]{@{}l@{}}Fine-tuned \\ End2end MLLMs\end{tabular}}
%& VTimeLLM (w/ finetune) & 24.4 & 9.0 & 20.5 & 9.2 & 5.6 & 3.8 \\
& VTimeLLM (w/ finetune) & 28.0 & 8.6 & 19.8 & 9.3 & 5.6 & 3.8 \\
%& \multicolumn{1}{l|}{\textbf{VideoNarrator (Ours)}} & \textbf{30.7} & \textbf{9.5} & \textbf{22.3} & \textbf{9.7} & \textbf{5.9}  & \textbf{4.1}  \\
& \multicolumn{1}{l|}{\textbf{VideoNarrator (Ours)}} & \textbf{33.3} & \textbf{9.1} & \textbf{21.0} & \textbf{9.9} & \textbf{6.3}  & \textbf{4.4}  \\
& \multicolumn{1}{l|}{a. w/o LLM LoRa finetune} & 10.5 & 6.9 & 16.5 & 5.4 & 2.8 & 1.7 \\ 
%& \multicolumn{1}{l|}{b. w/o Storyline} & 22.4 & 8.5 & 20.0 & 8.1 & 4.7 & 3.2 \\
& \multicolumn{1}{l|}{b. w/o Storyline} & 26.2 & 8.2 & 19.4 & 9.4 & 5.9 & 3.6 \\
%& \multicolumn{1}{l|}{c. w/o Video Clip Position} & 23.2 & 8.3 & 19.2 & 8.2 & 5.0 & 3.4 \\
& \multicolumn{1}{l|}{c. w/o Video Clip Position} & 29.5 & 8.7 & 20.1 & 9.1 & 5.6 & 3.9 \\
%& \multicolumn{1}{l|}{d. w/o Visual Compression} & 24.0 & 8.5 & 19.5 & 8.3 & 4.9 & 3.3 \\
& \multicolumn{1}{l|}{d. w/o Visual Compression} & 29.6 & 8.7 &20.4 & 9.2 & 5.7 & 3.8 \\
%& \multicolumn{1}{l|}{e. w/o Visual Memory} & 20.7 & 8.0 & 18.6 & 7.6 & 4.5 & 3.1 \\
& \multicolumn{1}{l|}{e. w/o Visual Memory} & 22.3 & 8.2 & 19.2 & 8.1 & 4.7 & 3.1 \\

\bottomrule
\end{tabular}
\end{table*}

\begin{table*}[t]
\caption{\label{table:auto_metric_2} Automatic Evaluation Results on E-SyncVidStory, considering four aspects (Section \ref{sec:metrics}): visual relevance, knowledge relevance, controllable accuracy, and fluency. The best result on each score is \textbf{bolded} and the second best is \underline{underlined}.
}
\vspace{-8pt}
\centering
\fontsize{7.2}{8.2}\selectfont
\begin{tabular}{l|cc|cc|c|c}
\toprule
 & \multicolumn{2}{c|}{\textbf{Visual Relevance}} & \multicolumn{2}{c|}{\textbf{Knowledge Relevance}} & \textbf{Controllable Acc.} & \multicolumn{1}{c}{\textbf{Fluency}} \\
 & \textbf{EMScore$\uparrow$} & \textbf{EMScore} $_\textrm{ref}\uparrow$ & \textbf{Info}$_\textrm{Sim}\uparrow$ & \textbf{Info}$_\textrm{Diverse}\uparrow$ & \textbf{Word Length$\uparrow$} & \multicolumn{1}{c}{\textbf{Intra-Repetition$\downarrow$}} \\ \midrule
LLaVA-1.5+GPT-3.5 (Zero-Shot) & 52.3 & 84.0 & 88.0 & 42.6 & 37.3 & 34.3 \\
%LLaVA1.5+GPT3.5 (One-Shot) & 53.4 & 84.5 & 84.6 &  & 48.0 &  \\ 
 LLaVA-1.5+GPT-3.5 (Few-Shot) &  52.6 &  84.3 &  \underline{88.4} &  46.9 &  50.7 &  23.2 \\
 \midrule
Video-ChatGPT \cite{maaz2023video_chatgpt} & 52.1&	72.7&	87.7&	39.2 &	17.4	&44.2\\
Video-LLaVA \cite{lin2023videollava}&  52.3	& 73.0 & 87.6 & 32.3 & 28.6	& 34.6\\
VTimeLLM \cite{huang2023vtimellm} & 52.1	& 82.9	&87.4	&40.2&	15.3&	41.6\\
\midrule
%VTimeLLM (w/ finetune) & 52.2 & 84.1 & 87.8 & 46.1 & 66.1 & 29.6\\
VTimeLLM (w/ finetune) & 52.4	&84.1	&88.0	& 46.8	& 70.2	& 25.7\\
%\textbf{VideoNarrator (Ours)} & \underline{52.9} & \textbf{85.3} & \textbf{88.8} & \textbf{55.2} & \textbf{98.2}& 10.8 \\
\textbf{VideoNarrator (Ours)} & \underline{52.8} & \textbf{85.1} & \textbf{88.6} & \textbf{50.2} & \textbf{98.1}& 10.8 \\
%a. w/o LLM LoRa finetune & 52.0 & 84.1 & 86.1 & 42.7 & 59.1 & \textbf{9.6} \\
a. w/o LLM LoRa finetune & 52.0 & 83.6 & 86.1 & 42.7 & 42.3 & \textbf{7.1} \\
%b. w/o Storyline & 52.5 & 85.0 & 83.3 & 52.2 & \underline{96.9} &  12.5\\
b. w/o Storyline & 52.3 & 84.2 & 88.3 & 49.2 & 96.3 &  \underline{8.6}\\
%c. w/o Video Clip Position & 52.6 & 84.5 & 87.7 & 49.9  & 96.0 & 13.6 \\
c. w/o Video Clip Position & 52.6 & 84.5 & 88.2 & 48.4 & 96.0 & 11.7 \\
%d. w/o Visual Compression & \textbf{53.0} & \underline{85.2} & 87.9 & \underline{52.3} & 93.3 & 11.3 \\
d. w/o Visual Compression & \textbf{52.9} & \underline{84.8} & 87.9 & \underline{50.0} & \underline{98.0} & 11.3 \\
e. w/o Visual Memory & 52.3 & 84.4 & 87.0 & 43.8  & 92.3 & 11.8 \\
\bottomrule
\end{tabular}
\end{table*}

\subsection{Prompting}
The input prompt for VideoNarrator is displayed in Figure \ref{fig:prompt}.
%When we try to input the overall video information and generate the entire story in one go,. Therefore, we opt for sequential generation instead. 
For each video clip, the controllable signal is provided just before the narration, which can ensure controllable accuracy. In our instruction, we concisely explain the task requirements. As shown in Figure  \ref{fig:prompt}, these red words can make it easier for the LLM to target relevant information. In the underlined instruction, we emphasize that the model should primarily use the information provided in the prompt and avoid using unrelated knowledge stored in the LLM.

\subsection{Training}
As shown in Figure \ref{fig:prompt}, we construct our dataset as multimodal instruction training samples. For a video consisting of $n$ clips, we maximize the probability of generating each script label $l_i$ and narration $s_i$ as follows:
$$
\sum_{i=1}^n P(l_i,s_i|X_\textrm{Instrution},\mathcal{K}, r_i, v_{j\le i} )
$$
We update the parameters of the video projector and the video clip positional embedding layer. Additionally, we apply LoRa \cite{hu2022lora} adapter to fine-tune the LLM, enabling it to possess the ability of controllable generation and relevant knowledge combination.
\begin{table}[t]
\caption{\label{table:with_storyline} Automatic evaluation results with predefined storyline. w.r.t CIDEr (C), EMScore (EMS), Info$_\textrm{Sim}$ (Sim), Info$_\textrm{Diverse}$ (Div), Word Length (Len), Script Label (Label), Intra-Repetition (IR).}
\vspace{-4pt}
\centering
\fontsize{5.4}{7.6}\selectfont
\begin{tabular}{l|c|c|cc|cc|c}
\toprule
 & \multicolumn{1}{c|}{\textbf{Text}} & \textbf{Visua}l & \multicolumn{2}{c|}{\textbf{Knowledge}} & \multicolumn{2}{c|}{\textbf{Controllable}} & \multicolumn{1}{c}{\textbf{Fluency}} \\
 & \multicolumn{1}{c|}{\textbf{C}} & \textbf{EMS$_\textrm{ref}$} & \textbf{Sim} & \textbf{Div} & \textbf{Len} & \textbf{Label} & \multicolumn{1}{c}{\textbf{IR}} \\ \hline
\textbf{VideoNarrator} & \textbf{40.1} & \textbf{86.1} & \textbf{88.2} & \textbf{53.1} & \textbf{98.8} & 95.4 &9.8 \\
a. w/o LoRa & 14.9 & 84.5 & 87.1 & 46.0 & 41.0 & 87.2 & \textbf{8.0}\\ 
c. w/o Pos. & 32.2 & 85.1 & 87.6 & 48.1 & 96.5 & \textbf{95.9} & 8.1 \\
d. w/o Comp. & 36.7 & 85.6  & 87.9 & 51.8 & 97.8 & 94.1 &  9.2\\
e. w/o Mem. & 29.5 & 85.3 & 87.6 & 49.0 & 92.3 & 93.1 &  8.3 \\\bottomrule
\end{tabular}
\end{table}
\vspace{-8pt}

\section{Experiments}

\begin{table*}[t]
\caption{\label{table:video_story} Automatic evaluation results on Video Storytelling \cite{li2019video} dataset.}
\centering
\fontsize{6.5}{8.2}\selectfont
\begin{tabular}{l|c|cccccc|c}
\toprule
 & Model Type & \textbf{METEOR} & \textbf{CIDEr} & \textbf{BLEU-1} & \textbf{BLEU-2} & \textbf{BLEU-3} & \textbf{BLEU-4} & \textbf{Length Acc.}\\
 \midrule
%VideoChat \cite{li2023videochat}& Generative & 15.49 & 42.9 & 50.00 & 43.30 & 34.76 & 27.21 \\
GVMF \cite{lu2022video} & Retrieval & 20.7 & 107.7 & 70.5 & 44.3 & 26.9 & 15.9 &- \\
VerbalizeVideos-GPT3.5 \cite{bhattacharya2023video} & Generative & 24.8 & 102.4 & 63.8 & 56.4 & 47.2 & 38.6 &-\\
VerbalizeVideos-BLIP2 \cite{bhattacharya2023video} & Generative &21.7 &108.9 & 55.2 & 48.5 & 40.7 & 33.8 &-\\
\textbf{VideoNarrator (Ours)} & Generative & \textbf{28.8} & \textbf{116.8} & \textbf{86.3} & \textbf{69.7} & \textbf{54.4} & \textbf{42.9} & 94.1\\
\bottomrule
\end{tabular}
\end{table*}

\subsection{Experiment Settings}
\paragraph{{Implementation Details.}} 
In our experiments, we use pre-trained models as visual feature extractors as described in the above Section. In experiments on E-SyncVidStory (advertising domain), we employ the Baichuan-7B model \cite{yang2023baichuan} as the LLM model, known for its excellent performance in Chinese-related tasks. We construct our dataset as illustrated in Figure \ref{fig:prompt}, resulting in 41k instruction samples. We apply 90\% of them as the training set and the left as the testing set. Our proposed framework is trained for 15 epochs, using a learning rate of $1e^{-4}$ and a batch size of 32. The LoRA parameters are set
to $r=64$ and $alpha=16$. The training of our 7B model took around $48$ hours on 4 A6000 GPUs.

In experiments on Video Storytelling dataset (general domain) \cite{li2019video}, the LLM model is replaced by LLaMA-2 \cite{touvron2023llama} for English generation. We make minor adjustments to fit the task setting of Synchronized Video Storytelling. Specifically, we apply the text length of the ground truth as the length requirement, and use the video topic as its relevant knowledge. The training prompt is similar to that presented in Figure \ref{fig:prompt}, while the system role is changed to ``You are an AI assistant that can understand videos'', the product information is replaced with the event topic, and the requirement for script labels has been removed.

%\subsection{Baselines}
%\noindent \textbf{Baselines}
%\paragraph{\textit{Compared Baseline: CaptioningModel+LLM}} We apply LLaVA \cite{liu2023visual} to get textual summaries for video clips and prompting GPT-3.5-Turbo-1106 \footnote{https://platform.openai.com/docs/models/gpt-3-5} for zero-shot and few-shot generation. Specifically, we select three example video stories with the highest click-through rate from the same domain and apply them in the prompt under the few-shot generation condition.

\paragraph{{Compared Baselines}} We compare our model with the following baselines. (1) \textit{Multi-model Pipieline.} Specifically, we apply LLaVA \cite{liu2023visual} to get textual summaries for video clips and prompting GPT-3.5 for zero-shot and few-shot generation. (2) \textit{End2end MLLMs}, including zero-shot performance of Video-ChatGPT \cite{maaz2023video_chatgpt}, Video-LLaVA \cite{lin2023videollava}, and VTimeLLM \cite{huang2023vtimellm} with 7B LLMs. These models take the video as inputs and generate narrations in an end2end manner. (3) \textit{Fine-tuned End2end MLLMs}. Specifically, we finetune the VTimeLLM model, which is effective for dense video comprehension and supports Chinese generation. More details about these baselines can be found in our Appendix \ref{sec:detaisl}.

As for experiments on the Video Storytelling dataset \cite{li2019video}, we compare our model with state-of-the-art methods: the retrieval-based GVMF \cite{lu2022video} and the generative-based VerbalizeVideos \cite{bhattacharya2023video}.

%\subsection{Ablation Studies}
\paragraph{{Ablation Study Setting.}} We carry out ablation studies to validate the contribution of different components in our proposed framework VideoNarrator. 

\noindent\textbf{a. \textit{w/o LLM LoRa Finetune}} Keep the LLM frozen during training in order to verify whether the LLM is effective for controllable generation.
%\paragraph

\noindent\textbf{b. \textit{w/o Storyline}} To verify whether the model can generate a nicely structured video story without provided explicit script labels. 
%While not apply explicit script labels to control the generation, whether the model can generate a nicely structured video story.

%In order to verify that Large Language Models are not effective for the controllable generation, the LLM is frozen during training.
%\paragraph{w/o Sequential Generation} In order to verify the effectiveness of our proposed structure.
\noindent\textbf{c. \textit{w/o Video Clip Position}} To verify the effectiveness of relative clip position embedding.

\noindent\textbf{d. \textit{w/o Visual Compression}} Retain all frames to verify the effectiveness of visual compression.

\noindent\textbf{e. \textit{w/o Visual Memory}}  Retain all video features of previous clips to verify the effectiveness of memory consolidation.%, video features of previous clips are all retained.

\begin{figure*}[t]
    \centering
    \includegraphics[width=\linewidth]{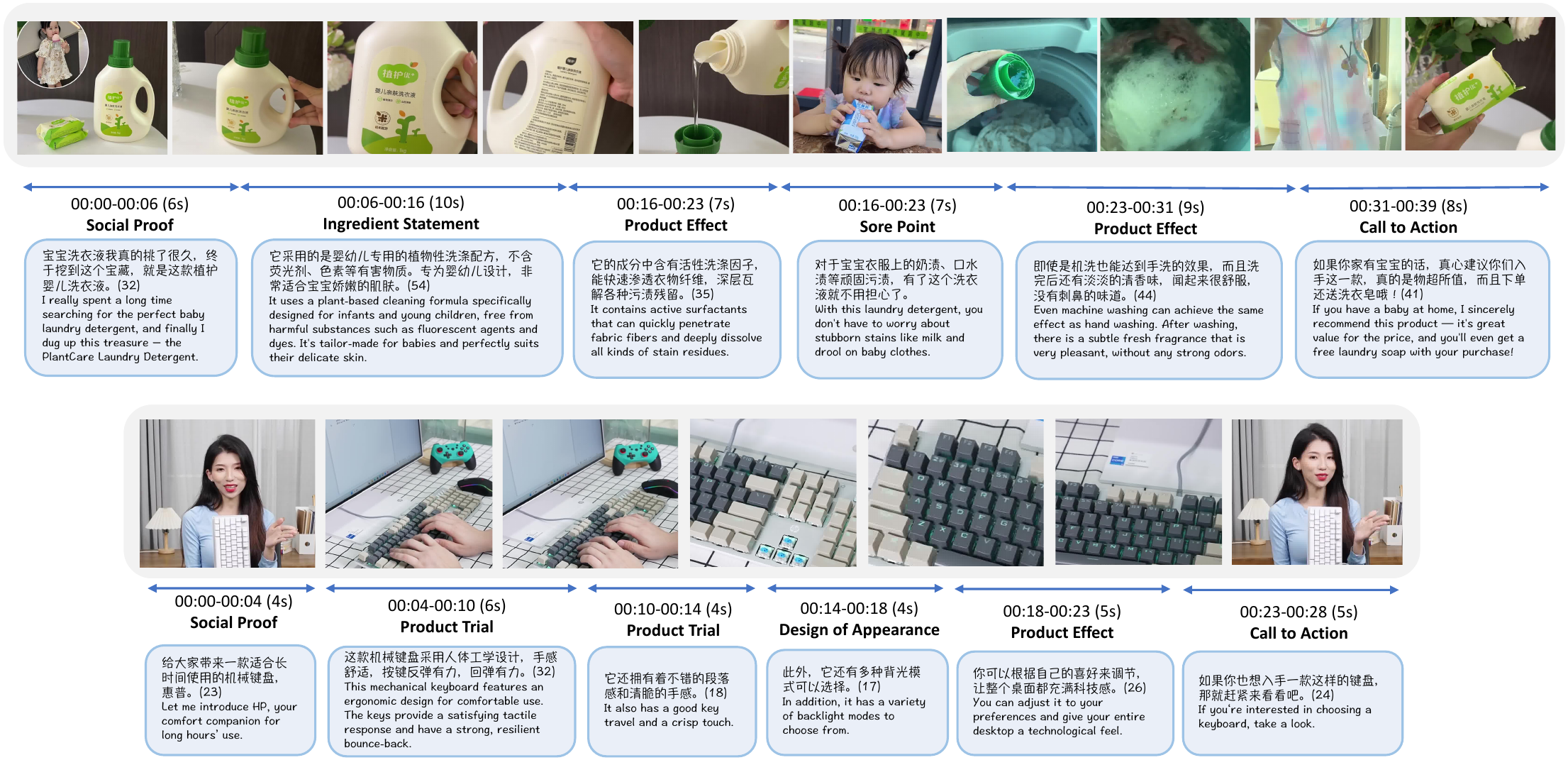}
    \vspace{-6pt}
    \caption{Examples generated by our proposed VideoNarrator.}
    
    \label{fig:case}
    \vspace{-4pt}
\end{figure*}
%\section{Results}

\subsection{Automatic Evaluation Results}
\paragraph{Advertising Video Storytelling.}
VideoNarrator can either generate a storyline to guide the generation, or use a predefined storyline to generate a story based on user preference. We conduct experiments for both situations on E-SyncVidStory. 
% compared with no label and gpt3.5
% ablation stud

\noindent\textit{\textbf{With Generated Storyline.}} As shown in Table \ref{table:auto_metric}, our model outperforms the baseline models, even when fine-tuned on our dataset. For detailed evaluations shown in Table \ref{table:auto_metric_2}, existing MLLMs demonstrate their ability to generate visually relevant and informative stories. However, they tend to focus on a limited amount of prior knowledge and may repeat it, resulting in a lower $\textrm{Info}_\textrm{Diverse}$ score. Furthermore, their fluency and ability to control the length of word sequences are poor. Based on the ablation studies, we find that by fine-tuning the LLM with our instruction data, our model can achieve the ability to control the word sequence length and comprehend video features, showing an increase in controllable accuracy and visual relevance. Applying visual compression and visual memory consolidation can improve the model's ability to extract relevant knowledge and retain story coherence, resulting in better performance in knowledge relevance and fluency. Explicit storylines and relative positional embedding can encourage narrations to follow a more logical outline of persuasion, 
improving the performance in knowledge relevance. 

\noindent\textit{\textbf{With Predefined Storyline.}}
Our model is also capable of generating stories with predefined storylines. Users can modify the generated storyline or offer a specific storyline to control the generation process, generating customized narrations. Table \ref{table:with_storyline} shows the results generated with the predefined storyline, demonstrating ideal performance in controllable accuracy and other aspects.

\paragraph{General Domain Video Storytelling}
To validate the effectiveness of our proposed VideoNarrator, we conduct experiments on the Video Storytelling dataset \cite{li2019video}, which covers a more general domain. As shown in Table \ref{table:video_story}, our model surpasses the state-of-the-art results. This showcases its effectiveness in synchronized video storytelling across a broader domain.

%Additionally, we made slight modifications to the existing Video Storytelling benchmark \cite{li2019video} to align with the task setting of Synchronized Video Storytelling. The results, as shown in  Appendix \ref{sec: general}, confirm the effectiveness and generalizability of our model in a more general domain.

\subsection{Human Evaluation and Qualitative Cases}

\begin{table}[t]
\centering
\fontsize{6.3}{8.9}\selectfont
\caption{\label{table:human_eval}Human evaluation results on three aspects. Geometric Mean (GM) demonstrates the overall performance.}
\vspace{-2pt}
\begin{tabular}{l|ccc|c}
\toprule
 & \textbf{Relevance}& \textbf{Attractiveness} & \textbf{Coherence}  & \textbf{GM}\\
\midrule
\multicolumn{1}{l|}{LLaVA1.5+GPT3.5} & 1.82 & 1.17 & 1.22 & 1.37\\
\multicolumn{1}{l|}{GPT4V+GPT4} & \textbf{2.48} & 2.05 & 2.18 & 2.22\\
\multicolumn{1}{l|}{\textbf{VideoNarrator}}& 2.34 & \textbf{2.32}   & \textbf{2.72} & \textbf{2.45} \\
\bottomrule
\end{tabular}
\vspace{-4pt}
\end{table}

We conduct human evaluations as well on three metrics: (1) \textit{Visual Relevance} measures the relationship between the generated narrations and the video shots. (2) \textit{Attractiveness} measures how well the stories can invoke the users' interests. (3) \textit{Coherence} measures the inter-sentence coherence and the completeness of the whole story. All metrics are rated from 0 to 3. We compare our models with the few-shot generation results of existing powerful LLMs. Specifically, we apply LLaVA1.5 or GPT-4V to generate descriptions for each visual scene. The captions of LLaVA1.5 are used as inputs for GPT3.5 to generate stories, and the ones generated by GPT-4V are provided for GPT4. 
The prompt is shown in Appendix \ref{sec:few-shot}. We randomly sample 30 stories and ask 13 advertisers to conduct the evaluation. As shown in Table \ref{table:human_eval}, our model outperforms GPT-3.5. We find that GPT3.5 often generates redundant statements and sometimes describes the visual scene instead of generating attractive narration. Regarding GPT-4V+GPT4, its performance is significantly improved compared to GPT3.5. However, our model still achieves better results in terms of story coherence and attractiveness, as well as very similar visual relevance scores. Additionally, it lacks length controllability, with an accuracy of 55.0\%, which impacts its practical use.

Figure \ref{fig:case} displays some example stories generated by our proposed framework.
% \usepackage{multirow}
% \usepackage{multirow}

%\subsection{Qualitative Examples}
%Figure \ref{fig:case} displays example stories generated by our proposed framework.
\section{Conclusion}
This paper introduces the new task of synchronized video storytelling, which aims to generate synchronous narrations for sequential video scenes. These narrations are guided by a structured storyline and can incorporate relevant knowledge, resulting in a coherent and informative story. The new task is more practical than existing tasks of dense video captioning and visual storytelling because the generated narrations can be directly combined with the visual scenes, creating an engaging video-format story.  We collect a benchmark dataset called E-SyncVidStory and introduce an effective framework named VideoNarrator. Both automatic and human evaluations verify the effectiveness of our proposed framework. 
\section*{Limitations}
In this work, our main focus is on generating synchronous narrations based on given videos. 
However, in real life, it is also important to transform unstructured video clips into well-structured ones. This aspect can be further explored in future research with the support of our proposed E-SyncVidStory. Additionally, our VideoNarrator model still faces the challenge of hallucination. For instance, most training samples show the application of face cream to the face after rubbing it on the hand. As a result, the model sometimes assumes that people will do the same when viewing someone rubbing the product on their hands. To address this issue, more constraints will be incorporated in future work.
\section*{Ethics Statement}
We acknowledge the Code of Ethics and Professional Conduct and strictly adhere to the rules throughout this research. There are two potential ethical issues with our work. The first pertains to the data source, while the second relates to the use of crowdsourcing services.
\paragraph{Data Source.} The datasets are collected from public advertisement videos on the e-commerce website \footnote{https://www.taobao.com/}. Given the copyright considerations, we will only release the URLs and features of the videos. Furthermore, our data source does not contain any information that identifies individuals by name or any offensive content.

\paragraph{Crowdsourcing Services.} After refining the data using the powerful GPT-4, we only need to conduct further checking of the annotations. We have hired 5 workers to review and correct any remaining errors. Each video took approximately 2 minutes to complete, and the workers received reasonable payment for the local area.
%and the workers were paid 1 RMB per video, which is a reasonable payment for local people.

\section*{Acknowledgements}
We thank all reviewers for their insightful comments and suggestions. 
This work was partially supported by the  National Natural Science Foundation of China (No. 62072462) and the Beijing Natural Science Foundation (No. L233008).

%\section*{Acknowledgements}

\bibliography{main}

\appendix

\section{Appendix}

\label{sec:appendix}

%\subsection{Experiments on General Domain Video Storytelling} \label{sec: general}
%To validate the effectiveness of our proposed VideoNarrator, we conduct experiments on the Video Storytelling dataset \cite{li2019video}, which covers a more general domain. We make minor adjustments to the dataset to fit the task setting of Synchronized Video Storytelling. Specifically, we apply the text length of the ground truth as the length requirement, and use the video topic as its relevant knowledge. The training prompt is similar to that presented in Figure \ref{fig:prompt}, while the system role is changed to ``You are an AI assistant that can understand videos'', the product information is replaced with the event topic, and the requirement for script labels has been removed. Additionally, the LLM model is replaced by LLaMA-2 \cite{touvron2023llama} for English generation. As shown in Table \ref{table:video_story}, our model surpasses the state-of-the-art results. This showcases its effectiveness in synchronized video storytelling across a broader domain.

\subsection{Details of Compared Baselines} \label{sec:detaisl}
\paragraph{Multi-model Pipelines} Following \citet{zhang2023video}, we convert the visual information of sequential video clips into textual summaries. This information, along with the product details and length requirements, is fed into a strong generative model GPT-3.5 to produce coherent narrations. Specifically, we apply key frames from each video clip (as described in Section \ref{sec:visual_compress}), generate English captions, and translate them into Chinese. We combine these captions to create the visual summaries. Figure \ref{fig:few-shot} shows our generation prompt. For few-shot generation, we select three example video stories with the highest click-through rate from the same domain and apply them in the prompt.
\begin{figure}[t]
    \centering
    \includegraphics[width=0.9\linewidth]{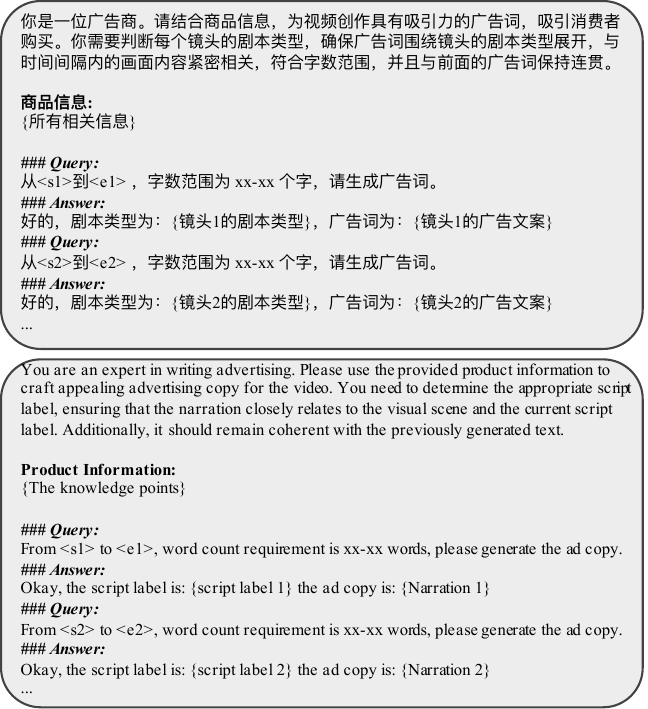}
    \caption{This figure illustrates the prompt template for VTimeLLM, with English translations provided for easy reading. }
    \label{fig:vtim}
\end{figure}

\begin{figure*}[ht]
    \centering
    \includegraphics[width=0.9\linewidth]{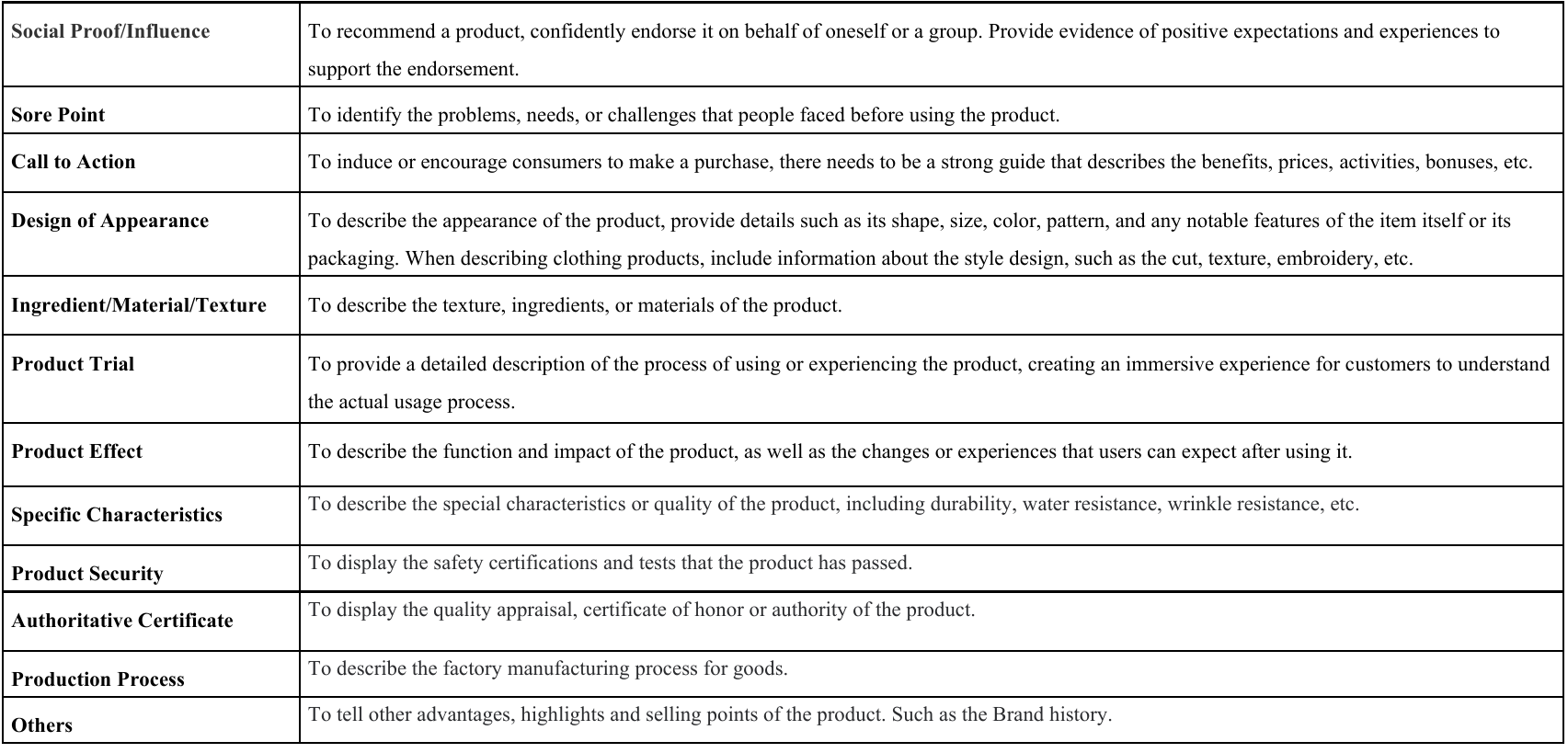}
    \caption{The definition of the script labels.}
    \label{fig:label_define}
\end{figure*}

\paragraph{End2end MLLMs} Most existing MLLMs, although promising, struggle to directly generate captions aligned correctly with sequential video clips. When an entire video is inputted, they often fail to properly ground the correct time ranges for each clip \cite{huang2023vtimellm}. For such models, we generate narrations sequentially. Specifically, we input the current video clip along with the previously generated narrations to create the current narration. We make experiments with Video-LLaVA \cite{lin2023videollava} and Video-ChatGPT \cite{maaz2023video_chatgpt}, which show SOTA results on video understanding tasks. Since these models are pre-trained in English only, we apply translations using the GPT-4 model. We then manually create well-designed instructions: ``You are a creative advertiser who is very familiar with the advertisement video.$\textbackslash n$ Product Information: \{knowledge points\}$\textbackslash n$ Previous advertisement copy: \{generated narrations\} $\textbackslash n$ Please proceed with the previous advertisement copy and create a new advertisement copy inspired by the video, using between \{x1\} to \{x2\} words. Ensure that the advertisement copy closely relates to the visual scene.$\textbackslash n$ Advertisement copy: ''

We also conduct experiments using VTimeLLM \cite{huang2023vtimellm}, a model pre-trained for time-aligned video understanding and is effective for dense video captioning. It supports Chinese generation with the ChatGLM3 \cite{du2022glm} structure. The generation prompt is shown in Figure \ref{fig:vtim}.

\paragraph{Finegrained End2end Large Multimodal Models} To the best of our knowledge, VTimeLLM is the most effective Chinese MLLM for sequential video comprehension. We fine-tune VTimeLLM on our proposed dataset, which is transformed into its QA format, as illustrated in Figure \ref{fig:vtim}. As illustrated in Table \ref{table:auto_metric} and Table \ref{table:auto_metric_2}, our model surpasses the fine-tuned results in all metrics.

\subsection{Evaluation Metric for Knowledge Relevance}\label{sec:eval_function}
As described in Section \ref{sec:knowledge_relevance}, we evaluate two aspects of knowledge relevance: information similarity and information diversity. To evaluate $\textrm{Info}_\textrm{Sim}$, we calculate the similarity of each knowledge point with the entire sentence $s_i$, selecting the maximum score as the coarse knowledge similarity %$\textrm{Sim}_{sen}$.  
We also calculate the maximum similarity of each word in $s_i$, and the average value is considered as the fine-grained knowledge similarity. $\textrm{Info}_\textrm{Sim}$ is calculated as the average of coarse and fine-grained similarity. To evaluate $\textrm{Info}_\textrm{Diverse}$, we check if a knowledge point $k_i$ has a similarity higher than 0.9 with a sentence or its segmented words. If this condition is met, we consider that the sentence includes knowledge point $k_i$. The $\textrm{Info}_\textrm{Diverse}$ score is calculated as the number of covered knowledge points divided by the video time duration. Given a story $S=\{s_i\}_{i=1}^n$ and its prior knowledge $\mathcal{K}$, the knowledge relevance scores are evaluated by:

\begin{small}
\[
    \begin{aligned}
        \textrm{Info}_\textrm{Sim}&= \frac{1}{2|s_i|}\sum_{si}\bigg(\mathop{\max}_{k\in \mathcal{K}}f_{k}^Tf_{s_{i}}+\\&\qquad \qquad \qquad \frac{1}{|W(s_i)|}\sum_{w\in W(s_i)}\mathop{\max}_{k\in \mathcal{K}}f_{k}^Tf_{w}\bigg)\\
        \textrm{Info}_\textrm{Diverse}&=\frac{1}{T}\bigg|\mathop{\cup}_{si} \big\{ k_t \in \mathcal{K} \big| \mathop{\max}_{w \in W(s_i)\cup s_i } f^T_{k_t}f_w > 0.9\big\}\bigg|,\\
    \end{aligned}
\]
\end{small}
where $W(s_i)$ represents all the words in sentence $s_i$, $T$ refers to the duration of the whole video. $f_{s_i}, f_{k}$, and $f_w$ refer to the normalized embeddings of sentence $s_i$, knowledge point $k$, and segmented word $w$, respectively.

\begin{figure*}[t]
    \centering
    \includegraphics[width=\linewidth]{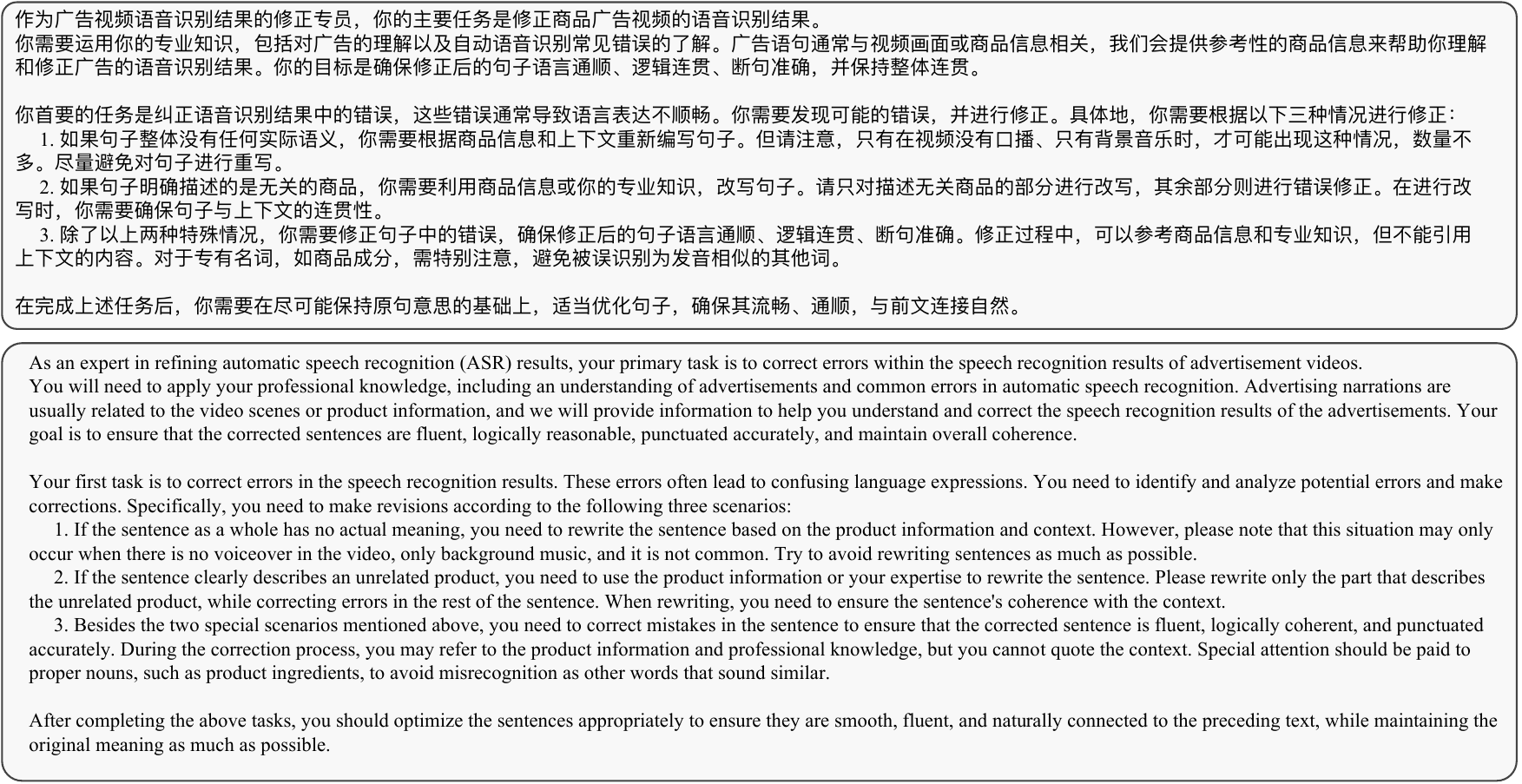}
    \caption{The prompt for automatic ASR refinement, with English Translation for easy reading.}
    \label{fig:asr}
\end{figure*}
\subsection{Human Evaluation Annotators}
All the annotators are employees from the advertising company and have a thorough understanding of advertisement videos. The team of 13 annotators includes 5 women and 8 men, ranging in age from 22 to 35. The Pearson Correlation score of their evaluation results is 0.55.

\subsection{Script Label Definition}\label{sec:scripts}
Script Labels are predefined structures to control the generated story. In this work, we focus on advertising stories in the e-commercial era. We analyze the collected advertising stories and categorize the script labels into 12 types: social proof/influence, sore point, call to action, design of appearance, ingredient/material/texture statement, product trial, product effect, product security, specific characteristics, authoritative certificate, production process, and others. Each of them corresponds to one of the advertising persuasion strategies \cite{kumar2023persuasion}. The definition of the 12 script labels is displayed in Figure \ref{fig:label_define}.%commonly used in the advertising domain.  

%\footnote{We analyze the collected advertising stories and categorize the script labels into 12 types as shown in Figure~\ref{fig:script_label}.
\begin{figure}[t]
    \centering
    \includegraphics[width=0.9\linewidth]{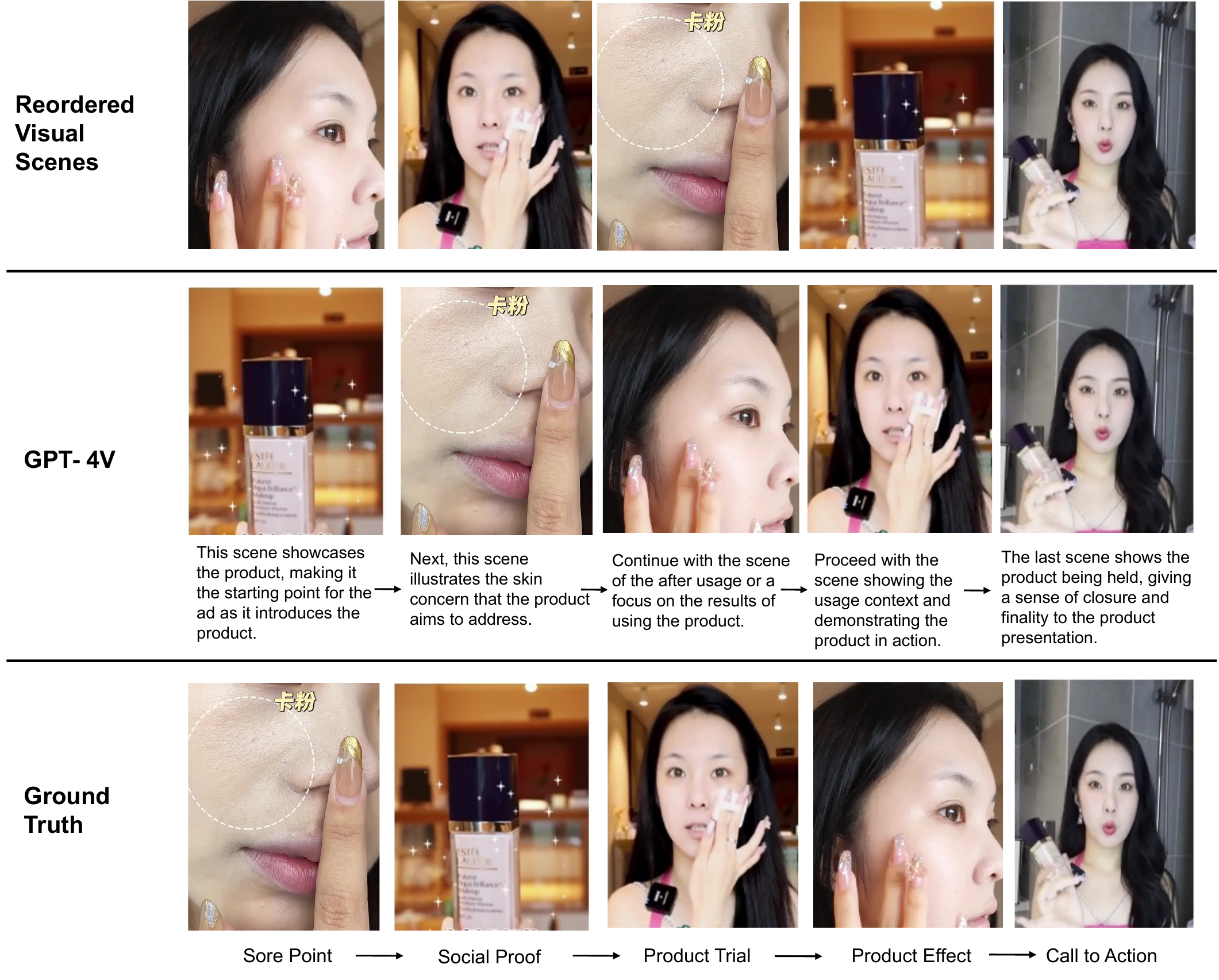}
    \caption{This figure illustrates the sorted results by GPT-4V.}
    \label{fig:resort}
\end{figure}
\begin{figure*}[t]
    \centering
    \includegraphics[width=\linewidth]{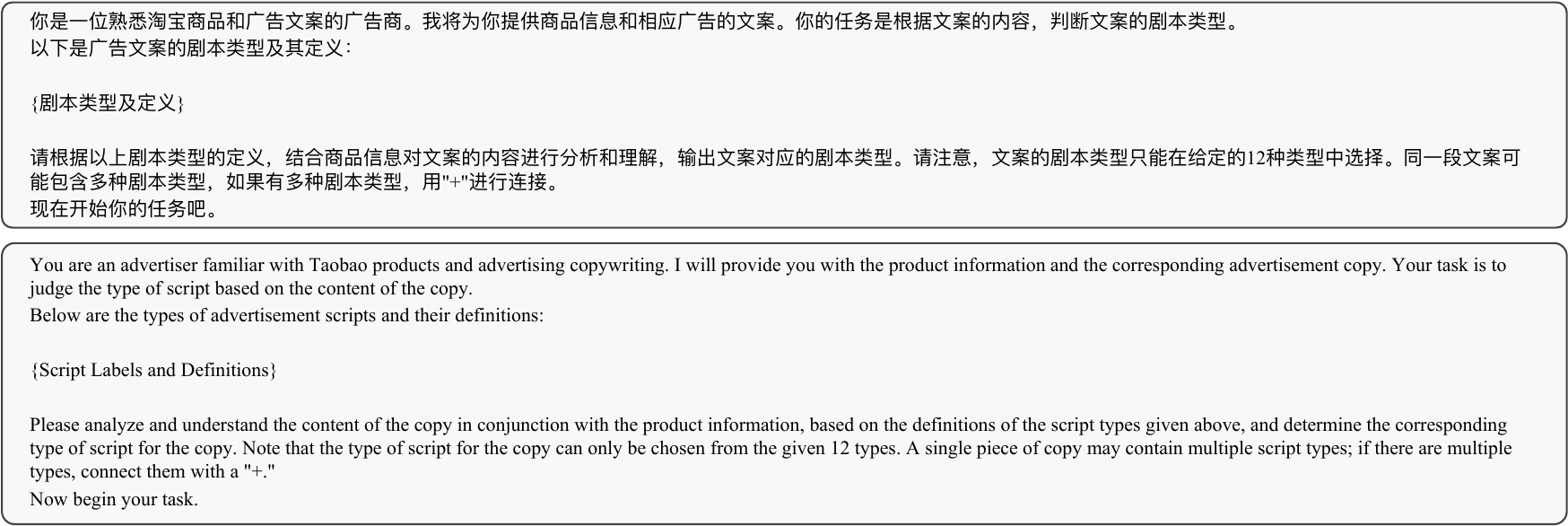}
    \caption{The prompt for script label classification, with English Translation for easy reading. The definition of script labels are detailed in Figure \ref{fig:label_define}}
    \label{fig:script_cls}
\end{figure*}

\subsection{Discussions of Further Research} \label{sec:beyond}
In our paper, the visual scenes are logically sorted. But at times, a user may wish to input unordered visual recordings and obtain sorted clips with a coherent narrative. We also hope to invoke additional research on the reordering task using the annotation provided in our proposed dataset. In Figure \ref{fig:resort}, we present the quantitative results sorted by the robust GPT-4V model. These results are still imperfect. Considering its reasoning process, the usage process should be positioned between the product's pre-use and post-use stages. Starting the advertisement with the pain point could make it more appealing, capturing the users' needs right from the beginning. This highlights the need for further research.

\subsection{Prompt for Data Processing}
\label{sec:prompt}

The prompt for ASR refinement is displayed in Figure \ref{fig:asr}. The prompt for script label classification is displayed in Figure \ref{fig:script_cls}.

\subsection{Prompt for Few-Shot Story Generation}\label{sec:few-shot}
\begin{figure*}[t]\vspace{-8pt}
    \centering
    \includegraphics[width=0.95\linewidth]{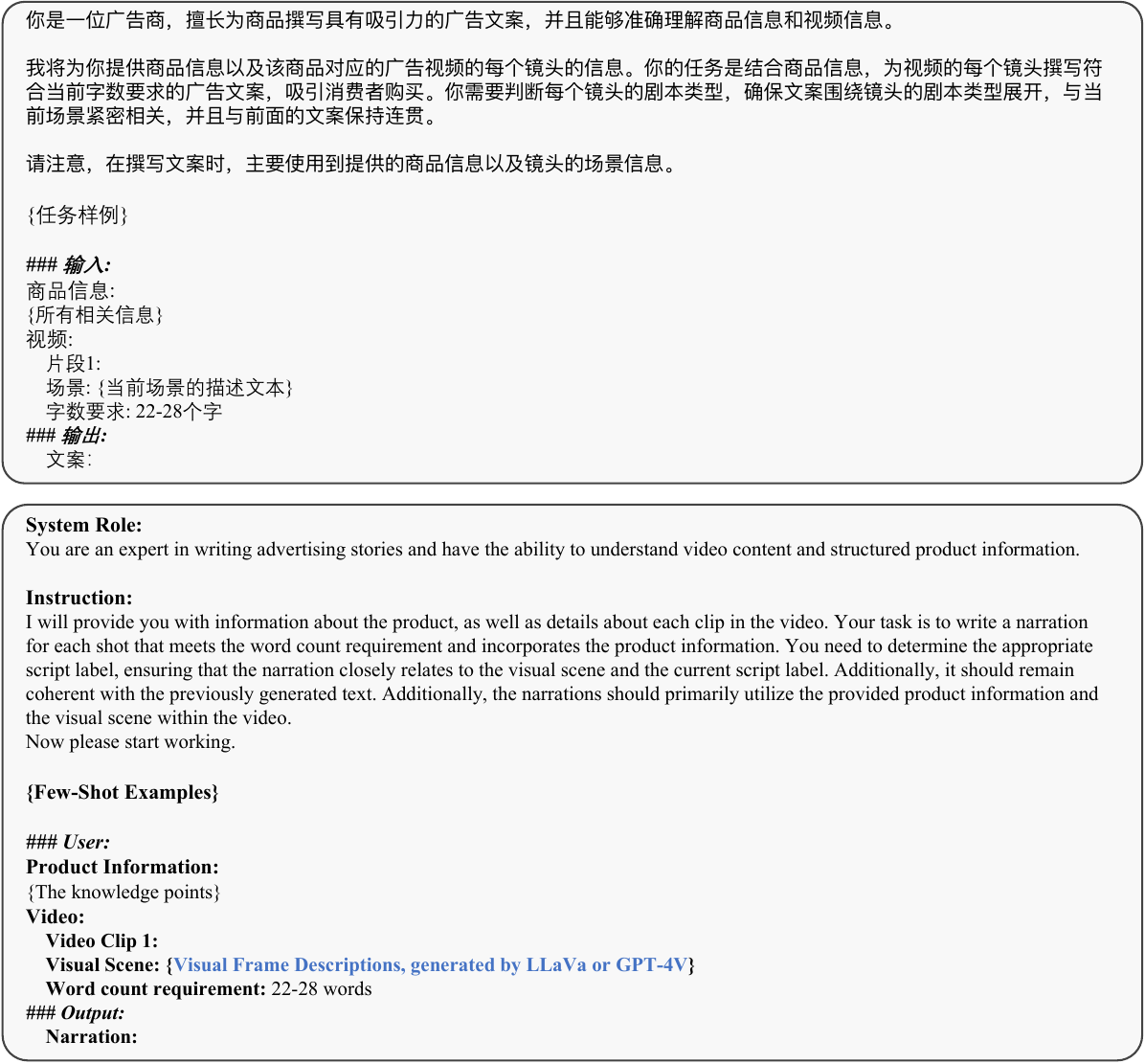}
\vspace{-4pt}
    \caption{The prompt for few-shot generation, with English Translation for easy reading.}
    \label{fig:few-shot}
\end{figure*}

The prompt for few-shot story generation with GPT3.5 and GPT4 is displayed in Figure \ref{fig:few-shot}.

\end{document}